\documentclass[pra,twocolumn,superscriptaddress,showpacs,amsmath,amssymb,amsfonts]{revtex4-1}
\usepackage[dvipdfmx]{graphicx,color}
\usepackage{subfigure}
\usepackage{bm}
\usepackage{siunitx}
\usepackage{amsthm,mathtools}
\usepackage[normalem]{ulem}
\usepackage{xcolor}

\def\U#1{{\rm #1}} 
\def\u#1{_{\rm #1}}
\newcommand{\ket}[1]{| #1 \rangle}
\newcommand{\braket}[2]{\langle #1 | #2 \rangle}
\newcommand{\vect}[1]{\boldsymbol{#1}}
\newcommand{\od}[2]{\frac{\mathrm{d} #1}{\mathrm{d} #2}}

\usepackage{lineno}

\begin{document}
\title{
  Loss-induced anomalous generalized bunching in multiphoton interference
}
\author{Rikizo Ikuta}
\email{ikuta.rikizo.es@osaka-u.ac.jp}
\affiliation{
  \mbox{Graduate School of Engineering Science, The University of Osaka,
  Osaka 560-8531, Japan}}
\affiliation{
  \mbox{Center for Quantum Information and Quantum Biology, The University of Osaka,
  Osaka 560-0043, Japan}}

\begin{abstract}
  We show that
  internal loss and survival conditioning can activate anomalous generalized bunching in passive linear optical circuits.
  We introduce a conditional bunching probability that all photons occupy a target region of accessible output modes, 
  given that all photons survive. 
  For two-photon inputs, we prove that this probability is always monotonic for any circuit size and loss configuration, 
  although a multimode target region can reverse the monotonic direction. 
  For three-photon inputs in a minimal three-mode lossy interferometer, 
  we find a nonmonotonic anomaly in which the conditional bunching probability is maximized
  for partially distinguishable photons. 
  This behavior is forbidden for the corresponding unconditioned target-region probability,
  demonstrating that survival-conditioned loss changes the minimal hierarchy of generalized bunching.
\end{abstract}

\maketitle

\section{Introduction}
Bosonic bunching is one of the most familiar signatures of quantum interference between identical particles.
In photonic systems, the simplest example is the Hong-Ou-Mandel~(HOM) effect~\cite{Hong1987}, 
where two identical photons incident on a balanced beamsplitter exit together from the same output port.
In the HOM setting, introducing distinguishability monotonically weakens the bunching effect
and restores coincidence events. 
This effect has shaped the standard intuition that increasing indistinguishability enhances 
the tendency of bosons to concentrate into fewer output modes. 
This intuition is related to the well-known $N!$ enhancement associated with Bose statistics, 
where the probability of finding identical bosons in the same output mode is enhanced
relative to that for distinguishable particles~\cite{FeynmanLecturesOnlineIII04}. 

\begin{figure}
 \begin{center}
      \scalebox{0.82}{\includegraphics{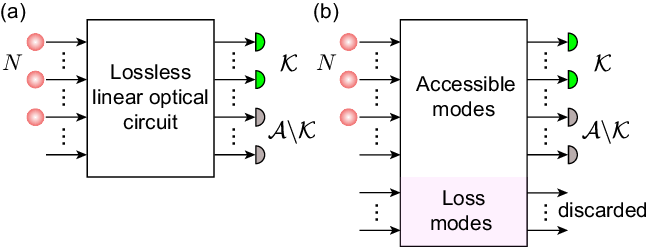}}
      \caption{
        (a) Standard generalized bunching setting. 
        (b) Generalized bunching setting considered in this paper,
        where the accessible modes are embedded in a larger mode space including loss modes.
        Only the accessible outputs in ${\mathcal A}$ are observed, while the loss outputs are discarded.
    }
 \label{fig:concept}
 \end{center}
\end{figure}

A natural formulation of this intuition is provided by generalized bunching~\cite{Shchesnovich2016,Seron2023},
as illustrated in Fig.~\ref{fig:concept}~(a). 
For $N$ photons injected into distinct input modes of a linear optical circuit, 
the generalized bunching probability is defined as the probability that all $N$ photons are detected 
within a prescribed target subset $\mathcal K$ of output modes. 
This definition includes ordinary bunching into a single output mode as a special case,
while extending the notion of bunching to concentration within a selected region of modes. 

The question of whether fully indistinguishable photons maximize generalized bunching 
is closely related to a conjecture on matrix permanents known as the permanent-on-top conjecture
and to recent studies of anomalous generalized bunching~\cite{Bapat1985,Bapat1986,Shchesnovich2016,Seron2023,Pioge2023,Pioge2026}. 
For $N=3$, the theorem in Ref.~\cite{Bapat1986} guarantees that the fully indistinguishable input 
maximizes the generalized bunching probability for any physically allowed internal-state Gram matrix of the input photons. 
However, the permanent-on-top conjecture itself is false in the broader matrix-theoretic setting for $N=5$~\cite{Shchesnovich2016}. 
A closely related permanent inequality~\cite{Bapat1985} was later shown to fail for $N=7$~\cite{Drury2016}. 
More directly relevant to the present linear-optical setting,
an explicit unitary circuit was provided in which the generalized bunching probability for $N=7$ photons
is maximized at partial distinguishability rather than at full indistinguishability~\cite{Seron2023}.

A related unconventional anti-bunching effect has also been reported in lossy two-photon interference at nonunitary beamsplitters~\cite{Vest2017,Hong2024}.
Both lines of work concern unconditioned output probabilities, 
leaving open how the bunching trend behaves after conditioning on photon survival. 
In this paper, we consider a survival-conditioned generalized-bunching setting,
as shown in Fig.~\ref{fig:concept}~(b). 
Instead of focusing on whether the fully indistinguishable input maximizes the unconditioned target-region probability, 
we ask how the target-region concentration behaves within the no-loss ensemble of a lossy linear optical circuit.
This conditional viewpoint is operationally natural in optical interference,
where loss can be represented as coupling to unobserved environmental modes
and normalized output statistics are often used to characterize interference among the surviving photons.

We show that, for $N=2$, the survival-conditioned probability is always monotonic.
For a single-mode target region, increasing indistinguishability enhances bunching, retaining the usual HOM-like trend. 
This result clarifies the distinction between raw lossy-HOM signatures~\cite{Vest2017,Hong2024} 
and bunching within the survival-conditioned ensemble. 
Multimode target regions can reverse the monotonic direction, 
so that fully distinguishable photons give the largest conditional bunching probability.
A genuinely nonmonotonic anomaly first becomes possible for $N=3$ in three accessible modes with a two-mode target region, where the conditional bunching probability can be larger at partial distinguishability than at full indistinguishability.
This anomaly persists after removing the output-transmission imbalance, 
indicating that it cannot be reduced to unequal output loss or detector efficiencies.
Thus, the survival-conditioned lossy setting bypasses the three-photon obstruction of the raw,
unconditioned generalized-bunching probability. 

\section{Standard generalized bunching}
We first review the standard generalized bunching in linear optical circuits~\cite{Shchesnovich2016,Seron2023}. 
As shown in Fig.~\ref{fig:concept}~(a),
consider $N$ single photons injected into $N$ distinct input modes of an $M$-mode unitary circuit $U$, with $N\leq M$. 
Let ${\mathcal K}\subseteq {\mathcal A} = \{ 1,\ldots,M \}$ be a prescribed target subset of the output modes. 
The generalized bunching probability is the probability that all $N$ photons are detected within ${\mathcal K}$. 
For partially distinguishable photons, the probability can be written as 
\begin{align}
P_{\mathcal K}^{(U)}(S)=\sum_{\vect{n}\in\Omega_{\mathcal K}}P_{\vect{n}}^{(U)}(S),
\label{eq:PKU}
\end{align}
where $\vect{n} = (n_1,\ldots,n_M)$ is an occupation-number tuple
specifying the output configuration satisfying $\sum_{j=1}^M n_j=N$, 
$P_{\vect{n}}^{(U)}(S)$ is the probability of observing the output pattern $\vect{n}$. 
$\Omega_{\mathcal K}$ denotes the set of all $N$-photon output configurations in which every occupied mode belongs to ${\mathcal K}$. 
$S$ is the Gram matrix of the internal states of the input photons and describes their pairwise distinguishability~\cite{Shchesnovich2015,Tichy2015}. 
By denoting the normalized internal state of the $i$th single photon by $\ket{\phi_i}$, 
the matrix elements of $S$ are given by 
\begin{align}
    S_{ii'}=\braket{\phi_i}{\phi_{i'}}.
\end{align}  

The naive bosonic bunching intuition suggests that fully indistinguishable photons should maximize the probability, namely $P_{\mathcal K}^{(U)}(S)\leq P_{\mathcal K}^{(U)}(S\u{id})$, 
where $(S\u{id})_{ii'}=1$ for all $i$ and $i'$. 
The anomaly of generalized bunching is the violation of this expectation. 
In Ref.~\cite{Seron2023}, it was shown that 
for $N=M=7$ with a suitably chosen unitary circuit $U$ and two-mode subset ${\mathcal K}$, 
there exists an internal-state Gram matrix $S$ such that 
\begin{align}
    P_{\mathcal K}^{(U)}(S) > P_{\mathcal K}^{(U)}(S\u{id}). 
\end{align}
This result established that, in sufficiently large unitary circuits, 
partial distinguishability can enhance the concentration of photons into a target region
beyond the fully indistinguishable limit. 

Importantly, in the standard unconditioned formulation, 
careful numerical searches suggested the absence of such anomalies for $N < 7$. 
In particular, the $N=3$ anomaly is ruled out by the permanent-on-top result 
even for a nonunitary transfer matrix~\cite{Bapat1986}. 
Thus, nonunitarity alone is insufficient to produce a three-photon anomaly 
at the level of the raw target-region probability. 
This obstruction, however, does not directly apply to the survival-conditioned ratio considered below.

\section{Generalized bunching in lossy optical circuits} 
We extend generalized bunching to lossy linear optical circuits.
As shown in Fig.~\ref{fig:concept}~(b),
loss is represented by coupling the accessible modes to unobserved environmental modes. 
Only the accessible output modes are measured, while the loss modes are discarded. 
The corresponding accessible transfer matrix is generally nonunitary,
so that conditioning on photon survival leads to a quantity distinct from the raw target-region probability. 

Formally, we consider $M$ accessible modes and $L$ loss modes. 
Let ${\mathcal A}=\{1,\ldots,M\}$ denote the set of accessible output modes,
and let ${\mathcal K}\subseteq {\mathcal A}$ be a target subset. 
The full circuit is described by an $(M+L)$-mode unitary operator $\hat{\mathcal U}$,
while only the modes in ${\mathcal A}$ are measured. 
The $N$ photons are injected one by one into $N$ distinct accessible input modes, with $N\leq M$. 
Without loss of generality, we label the occupied input modes as $i=1,\ldots,N$. 
All loss input modes are initially in the vacuum state. 
We assume that the unitary transformation acts only on the spatial modes without changing the internal states. 
For the $i$th input mode,
the creation operator $\hat{a}^{\dagger}_{i}(\phi_i)$ transforms as 
\begin{align}
  \hat{\mathcal U} \hat{a}^{\dagger}_{i}(\phi_i) \hat{\mathcal U}^\dagger
  =
  \sum_{j=1}^M T_{ji}\hat{b}^{\dagger}_j(\phi_i)
  +
  \sum_{\ell = 1}^{L}
  R_{\ell i}\hat{e}^{\dagger}_{\ell}(\phi_i),
\end{align}
where $\hat b_j^\dagger(\phi_i)$ and $\hat e_{\ell}^\dagger(\phi_i)$ are the creation operators for photons 
carrying the internal-state label $\phi_i$ in the $j$th accessible output mode and the $\ell$th loss mode, respectively. 
$T$ and $R$ are the $M\times N$ and $L\times N$ transfer matrices, respectively. 
For a fixed internal-state label $\phi$, the input and output spatial-mode operators satisfy
$[\hat{a}_{i}(\phi),\hat{a}^{\dagger}_{i'}(\phi)]=\delta_{ii'}$,
$[\hat{b}_{j}(\phi),\hat{b}^{\dagger}_{j'}(\phi)]=\delta_{jj'}$,
$[\hat{e}_{\ell}(\phi),\hat{e}^{\dagger}_{\ell'}(\phi)]=\delta_{\ell\ell'}$ and
$[\hat{b}_{j}(\phi),\hat{e}^{\dagger}_{\ell}(\phi)]=0$. 
Using the commutation relations and the unitarity of $\hat{\mathcal U}$,
we obtain $T^\dagger T + R^\dagger R = I_N$. 
Thus, the accessible transformation matrix $T$ is generally nonunitary as
\begin{align}
  T^\dagger T \leq I_N.
  \label{eq:TT}
\end{align}

Following Eq.~(\ref{eq:PKU}), we define the raw lossy generalized bunching probability as
\begin{align}
  P_{\mathcal K}^{(T)}(S) = \sum_{\vect{n}\in\Omega_{\mathcal K}} P_{\vect{n}}^{(T)}(S),
  \label{eq:PKT}
\end{align}
where $\vect{n}=(n_1,\ldots,n_M)$ satisfying $\sum_{j=1}^{M}n_j=N$. 
Similarly, the survival probability is obtained by summing over all accessible output patterns as 
\begin{align}
  P_{\mathcal A}^{(T)}(S) =  \sum_{\vect{n}\in\Omega_{\mathcal A}} P_{\vect{n}}^{(T)}(S),
  \label{eq:PKA}
\end{align}
where $\Omega_A$ is the set of all $N$-photon output patterns over the accessible modes.
Importantly, $P_{\mathcal K}^{(T)}(S)$ is an unnormalized target-region probability.
It depends both on the probability that all $N$ photons survive 
and on how strongly the surviving photons are concentrated within the target region $\mathcal K$. 
To isolate the bunching structure within the no-loss ensemble, 
we introduce the normalized target-region probability as 
\begin{align}
  B_{\mathcal K}^{(T)}(S)  = \frac{P_{\mathcal K}^{(T)}(S)}{P_{\mathcal A}^{(T)}(S)}, 
  \label{eq:BK}
\end{align}
which is physically meaningful whenever $P_{\mathcal A}^{(T)}(S) > 0$. 
In the absence of loss, the accessible transformation becomes unitary as $T=U$, resulting in $P_{\mathcal A}^{(T)}(S)=1$. 
Consequently, $B_K^{(T)}(S)$ reduces to the ordinary generalized bunching probability in Eq.~(\ref{eq:PKU}).

In discussing optical loss related to Eq.~(\ref{eq:TT}),
it is important to distinguish losses outside the interferometer from loss inside the interferometer.
External linear loss before the interferometer does not affect $B_{\mathcal K}^{(T)}(S)$,
because it contributes only a common multiplicative factor to all surviving $N$-photon events. 
In contrast, by reweighting different accessible output patterns, 
mode-dependent output loss can produce a distinguishability dependence of $B_{\mathcal K}^{(T)}(S)$,
even though the internal interference structure is unchanged.
This is an experimentally relevant issue, for example when photon detectors have unequal quantum efficiencies.

Including the final mode-dependent output loss, the total accessible transfer matrix can be decomposed as 
\begin{align}
  T\u{ext} = D_{\eta} T\u{int},
  \label{eq:DT}
\end{align}
where $D_{\eta} = \U{diag}(\sqrt{\eta_1},\ldots,\sqrt{\eta_M})$ 
with $\eta_j$ denoting the efficiency associated with the $j$th accessible output mode, and 
$T\u{int}$ denotes the transfer matrix of the interferometer before the final output loss. 
The probability observed after the final output loss is related to the probability before that loss by 
\begin{align}
    P_{\vect{n}}^{(T\u{ext})}(S)
    =
    \left(
    \prod_{j=1}^{M}
    \eta_j^{n_j}
    \right)
    P_{\vect{n}}^{(T\u{int})}(S).
\end{align}
When we assume that the efficiencies $\eta_j$ can be characterized independently of $T\u{int}$, 
their effect can be removed by dividing each observed output probability by the corresponding factor 
$\prod_j\eta_j^{n_j}$ and renormalizing the resulting distribution.

As a complementary normalization,
we can remove the directly observable row weights of the total accessible transfer matrix. 
Defining
\begin{align}
  q_j = \sum_{i=1}^{N} \left| (T\u{ext})_{ji} \right|^2 = \eta_j \sum_{i=1}^{N} \left| (T\u{int})_{ji} \right|^2,
  \label{eq:qj}
\end{align}
where $q_j$ is the sum of the single-photon detection probabilities at the $j$th accessible output mode,
obtained by probing the input modes individually.
After omitting any output modes with $q_j=0$, we introduce the row-normalized transfer matrix as 
\begin{align}
  \overline{T}\u{int} = D_q^{-1}T\u{ext}, 
\end{align}
where $D_q = \U{diag} (\sqrt{q_1},\ldots,\sqrt{q_M})$. 
This procedure equalizes the total single-photon transmission of the accessible output modes
using the directly measured quantities $q_j$,
without requiring the final efficiencies $\eta_j$ to be separated from the output-mode weights of $T\u{int}$.
Since the observed probability for $\vect{n}$ satisfies 
\begin{align}
  P_{\vect{n}}^{(T\u{ext})}(S) =
  \left( \prod_{j=1}^{M} q_j^{n_j} \right) P_{\vect{n}}^{(\overline{T}\u{int})}(S), 
\end{align}
the probability distribution $P_{\vect{n}}^{(\overline{T}\u{int})}(S)$ 
can be obtained directly in postprocessing by dividing each observed $N$-photon probability
by $\prod_{j=1}^{M}q_j^{n_j}$ and renormalizing over all accessible $N$-photon output patterns.
We note that $\overline{T}\u{int}$ does not necessarily satisfy Eq.~(\ref{eq:TT}),
but this condition can always be met by an appropriate global rescaling of the matrix.
The conditional probability in Eq.~(\ref{eq:BK}) remains unchanged under such rescaling.

By comparing $B_K^{(T\u{int})}(S)$ and $B_K^{(\overline{T}\u{int})}(S)$, 
we can determine whether the anomalous distinguishability dependence persists
after removing the output-transmission imbalance quantified by the measured row weights $q_j$. 
An anomaly in $B_{\mathcal K}^{(T\u{int})}(S)$ cannot be attributed to external output loss
or detector-efficiency imbalance, since such effects are separated into $D_\eta$. 
If the anomaly also persists in $B_{\mathcal K}^{(\overline{T}\u{int})}(S)$,
it cannot be explained simply by a larger total single-photon transmission of the target output modes. 
Such persistence indicates that the effect is sensitive to the internal nonunitary interference structure,
rather than to output-transmission imbalance alone.

\section{Purpose of the study}
We specify our setting. 
We use the conditional generalized bunching probability defined above
together with the following one-parameter internal-state Gram matrix as 
\begin{align}
(S(x))_{ii'} = x + (1-x)\delta_{ii'}
\label{eq:Sx}
\end{align}
for $0\leq x\leq 1$. 
Here, $x=0$ and $x=1$ correspond to fully distinguishable and fully indistinguishable photons, respectively. 
For brevity, we write $B_{\mathcal K}^{(T)}(x)$ for $B_{\mathcal K}^{(T)}(S(x))$ when no confusion arises. 
Our aim is to characterize how the conditional generalized bunching probability
depends on $x$ in lossy linear optical circuits. 
In particular,
we ask for the smallest lossy setting in which $B_{\mathcal K}^{(T)}(x)$ can be maximized at partial distinguishability for $0 < x < 1$. 

Before turning to specific photon numbers,
we rewrite the target-region and survival probabilities
in Eqs.~(\ref{eq:PKT}) and (\ref{eq:PKA}) in a common permanent form, 
following the standard formulation of generalized bunching~\cite{Shchesnovich2016,Seron2023}. 
We define the positive semidefinite Gram matrix $H_{\mathcal R}$ by 
\begin{align}
  (H_{\mathcal R})_{ii'} = \sum_{j\in {\mathcal R}} T_{ji}^*T_{ji'}
  \label{eq:HR}
\end{align}
for ${\mathcal R}={\mathcal K}, {\mathcal A}$. 
The diagonal elements $(H_{\mathcal R})_{ii}$ give the single-photon probabilities
for input mode $i$ to reach the output region ${\mathcal R}$, 
while the off-diagonal elements $(H_{\mathcal R})_{ii'}$ describe
the corresponding interferometric cross terms between inputs $i$ and $i'~(\neq i)$. 
The probability $P_{\mathcal R}^{(T)}(S)$ is
\begin{align}
  P_{\mathcal R}^{(T)}(S) = \U{perm} \left( H_{\mathcal R}\odot S \right),
  \label{eq:perm}
\end{align}
where $\odot$ denotes the Hadamard product. 
This expression follows by summing the standard partially distinguishable output probabilities over all output configurations 
supported within $\mathcal R$.

\section{Two-photon conditional generalized bunching}
We consider two photons~($N=2$) propagating through an arbitrary number $M$ of accessible output modes.
From Eqs.~(\ref{eq:Sx}) and (\ref{eq:perm}), the probability $P_{\mathcal R}^{(T)}(x)$ is given by
\begin{align}
  P_{\mathcal R}^{(T)}(x) = a_{\mathcal R} + b_{\mathcal R}x^2, 
\label{eq:PK2}
\end{align}
where $a_{\mathcal R} = (H_{\mathcal R})_{11}(H_{\mathcal R})_{22}$ and $b_{\mathcal R} = \left|(H_{\mathcal R})_{12}\right|^2$. 
Therefore, the conditional generalized bunching probability in Eq.~(\ref{eq:BK}) is 
\begin{align}
  B_{\mathcal K}^{(T)}(x) = \frac{a_{\mathcal K}+b_{\mathcal K} x^2}{a_{\mathcal A}+b_{\mathcal A} x^2}. 
\label{eq:BK2}
\end{align}
The derivative with respect to $y=x^2$ becomes 
\begin{align}
  \od{B_{\mathcal K}^{(T)}(y)}{y} = \frac{b_{\mathcal K}a_{\mathcal A} - a_{\mathcal K}b_{\mathcal A}}{(a_{\mathcal A}+b_{\mathcal A} y)^2}.
  \label{eq:dBK2}
\end{align}
Thus, in the two-photon case,
$B_K^{(T)}(x)$ is monotonic or constant for any accessible transfer matrix $T$ with nonzero survival probability.
This result covers both $T\u{int}$ and $\overline{T}\u{int}$,
irrespective of the numbers $M$ and $L$ of accessible and loss modes, respectively.
We note that the same conclusion holds for a general complex two-photon overlap $s$,
since Eqs.~(\ref{eq:PK2}) -- (\ref{eq:dBK2}) remain valid after replacing $x^2$ with $|s|^2$. 

In particular, when the target region ${\mathcal K}$ consists of a single output mode~($|{\mathcal K}|=1$), 
$H_{\mathcal K}$ in Eq.~(\ref{eq:HR}) is a rank-one matrix and therefore satisfies
$\det(H_{\mathcal K})=a_{\mathcal K}-b_{\mathcal K} = 0$. 
Since $\det(H_A)=a_{\mathcal A}-b_{\mathcal A}\geq 0$ is satisfied, Eq.~(\ref{eq:dBK2}) shows that 
$B_{\mathcal K}^{(T)}(x)$ is a monotonically nondecreasing function of the photon indistinguishability 
and is maximized in the fully indistinguishable case of $x=1$. 
This means that neither internal nor external loss can reverse the conditional bunching trend.
For a two-output circuit, the coincidence probability conditioned on two-photon survival is given by 
$1- B_{\{1\}}^{(T)}(x) - B_{\{2\}}^{(T)}(x)$
and is therefore nonincreasing in $x$.
This result clarifies the distinction between raw lossy-HOM signatures~\cite{Vest2017,Hong2024}
and bunching within the survival-conditioned ensemble.

Next, we consider the case where the target region ${\mathcal K}$ contains more than one output mode. 
The smallest nontrivial setting is $M=3$ with a two-mode target region, for example ${\mathcal K} = \{1,2\}$. 
In this case, we give an explicit lossy optical circuit satisfying $B_{\mathcal K}^{(T)}(0) > B_{\mathcal K}^{(T)}(1)$
which demonstrates a reversal of the conditional bunching trend.

\begin{figure}
 \begin{center}
      \scalebox{1}{\includegraphics{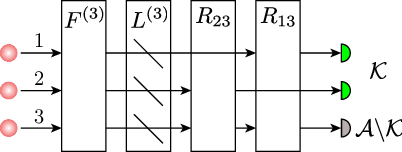}}
      \caption{
        The circuit in Eq.~(\ref{eq:circuit}) with $N=M=3$ and $K=\{ 1,2\}$.
        The diagonal loss $L^{(3)}$ is placed inside the interferometer,
        before the final beamsplitter stages.
        For the $N=2$ case, the third input mode is left empty.
    }
 \label{fig:example}
 \end{center}
\end{figure}
We consider the internal transfer matrix
\begin{align}
  T\u{int} = R_{13}(\theta_{13}) R_{23}(\theta_{23}) L^{(3)} F^{(3)}.
  \label{eq:circuit}
\end{align}
Here $(F^{(3)})_{jj'} = e^{2i \pi (j-1)(j'-1)/3}/\sqrt{3}$ is the three-mode discrete Fourier transform,
and $L^{(3)}=\U{diag}(\sqrt{\eta_1},\sqrt{\eta_2},1)$ is a mode-dependent loss matrix. 
$R_{ij}(\theta_{ij})$ denotes a beamsplitter matrix with real elements, acting as
\begin{align}
  R_{ij}(\theta_{ij})=
  \begin{pmatrix*}[r]
    \cos\theta_{ij} & \sin\theta_{ij}\\
    -\sin\theta_{ij} & \cos\theta_{ij}
  \end{pmatrix*}
\end{align}
on modes $i$ and $j$ and as the identity on the remaining mode.
Because $L^{(3)}$ is sandwiched between the initial Fourier transform and the final beamsplitter rotations,
it is not equivalent to a simple input or output loss.

With the parameters
$\eta_1 = \eta_2 = 1/3$ and $\theta_{13}=\theta_{23}=\pi/8$, 
we obtain 
$B_{\mathcal K}^{(T\u{int})}(0) \approx 0.283 > B_{\mathcal K}^{(T\u{int})}(1) \approx 0.258$ and 
$B_{\mathcal K}^{(\overline{T}\u{int})}(0) \approx 0.349 > B_{\mathcal K}^{(\overline{T}\u{int})}(1) \approx 0.325$. 
By contrast, for the corresponding $3 \times 2$ transfer matrix $T_0$ obtained from $L^{(3)}F^{(3)}$
by leaving the third input mode empty, we obtain
$B_{\mathcal K}^{(T_0)}(0)\approx 0.160 < B_{\mathcal K}^{(T_0)}(1) \approx 0.172$.
Its row-normalized matrix $\overline{T}_0$, which is proportional to the corresponding two-column submatrix of $F^{(3)}$,
also follows the conventional trend as $B_{\mathcal K}^{(\overline{T}_0)}(0) = 4/9 < B_{\mathcal K}^{(\overline{T}_0)}(1) = 5/9$.
Thus, the reversal cannot be attributed merely to output-row-weight imbalance. 

\section{Three-photon conditional generalized bunching}
We increase the input photon number to $N=3$.
The smallest accessible mode number is then $M=3$. 
We first consider the smallest target region $|{\mathcal K}|=1$. 
In this case, the conventional trend is recovered for any number $M\geq 3$ of accessible output modes, 
as proved in Appendix~\ref{appA}. 
We therefore proceed to the next minimal target-region size of $|\mathcal K|=2$,
in the minimal accessible-mode setting $M=3$, 
where anomalous generalized bunching appears.

\begin{figure}
 \begin{center}
      \scalebox{0.7}{\includegraphics{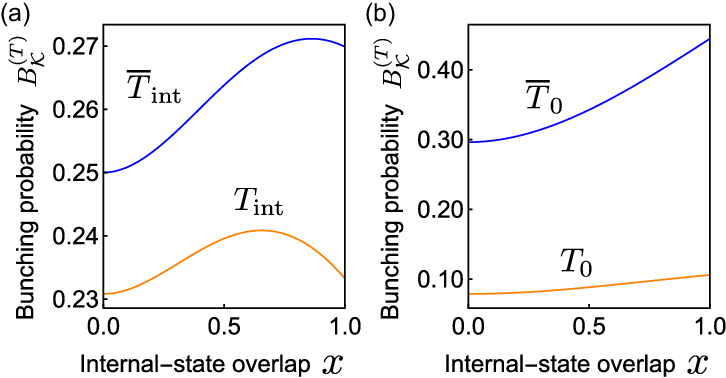}}
    \caption{
      (a) Conditional bunching probability for the full circuit 
      $T\u{int}=R_{13}(\pi/8)R_{23}(\pi/4)L^{(3)}F^{(3)}$.
      (b) For the circuit $T_0=L^{(3)}F^{(3)}$ without the final beamsplitter rotations. 
      The vertical scales are chosen independently in the two panels to make the variation with $x$ visible. 
    }
 \label{fig:anomaly}
 \end{center}
\end{figure}
We consider the case where ${\mathcal K}=\{ 1,2\}$, 
using the same circuit architecture as in Eq.~(\ref{eq:circuit}) and Fig.~\ref{fig:example},
but now inject one photon into each of the three input modes. 
For the internal loss matrix $L^{(3)}$, we take $\eta_1=1/2$ and $\eta_2=1/4$, and for the beamsplitter matrices, 
we set $\theta_{13}=\pi/8$ and $\theta_{23}=\pi/4$.
The resulting conditional bunching probabilities are shown in Fig.~\ref{fig:anomaly}~(a).
For the full circuit,
both $B_{\mathcal K}^{(T\u{int})}(x)$ and $B_{\mathcal K}^{(\overline{T}\u{int})}(x)$ exhibit a maximum at partial distinguishability. 
In particular, we find
$B_{\mathcal K}^{(T\u{int})}(x_{\max})\approx 0.241 > B_{\mathcal K}^{(T\u{int})}(1)\approx 0.233$ 
for $x_{\max}\approx 0.655$, and 
$B_{\mathcal K}^{(\overline{T}\u{int})}(\overline{x}_{\max})\approx 0.271 > B_{\mathcal K}^{(\overline{T}\u{int})}(1)\approx 0.270$ 
for $\overline{x}_{\max}\approx 0.862$. 
By contrast, when the final beamsplitter rotations are removed, both $T_0=L^{(3)}F^{(3)}$ 
and $\overline{T}_0=F^{(3)}$ are maximized at $x=1$, as shown in Fig.~\ref{fig:anomaly}~(b). 
Thus, for the present parameter set, 
the nonmonotonic maximum appears only after the final beamsplitter rotations and remains after row normalization.
For $N=M=3$, any transformation of the form $T=D_\eta U$, 
where $U$ is unitary and all output efficiencies are nonzero, 
is reduced to $\overline{T}=U$ by row normalization.
Hence, although mode-dependent output attenuation can produce an anomaly 
before row normalization for suitable parameters as shown in Appendix~\ref{appB}, 
such an anomaly necessarily disappears once the output-row imbalance is removed.
By contrast, the anomaly found here persists after row normalization,
showing that it cannot be reduced to output-mode attenuation 
and instead reflects the internal nonunitary interference structure.
The anomaly is not restricted to a fine-tuned parameter set but occurs over finite regions of the final beamsplitter angles,
both before and after row normalization, as shown by the parameter scans in Appendix~\ref{appC}. 

\section{Origin of the anomaly}
\begin{table}[t]
  \centering
\begin{tabular}{cccc}
    \hline
    $N$ & $|\mathcal K|$ & $M$      & Behavior of $B_{\mathcal K}^{(T)}(x)$ \\ \hline
    $2$ & 1              & $\geq 2$ & conventional trend \\
    $2$ & $\geq 2$       & $\geq 3$ & trend reversal possible \\
    $3$ & $1$            & $\geq 3$ & conventional trend \\
    $3$ & $2$            & $3$      & interior maximum or trend reversal possible \\
    \hline
  \end{tabular}
\caption{
  Summary of the results obtained in this work 
  for the one-parameter internal-state Gram matrix $S(x)$.
  The trivial case ${\mathcal K} = {\mathcal A}$, for which $B_{\mathcal K}^{(T)}(x)=1$, is excluded.
  }
  \label{tbl:result}
\end{table}
In the following, we discuss the origin of the three-photon anomaly from the weighting of permutation classes by partial distinguishability. 
For two photons~($N=2$), only the identity and pairwise-exchange classes contribute,
leading to a ratio of two linear functions of $x^2$ as seen in Eq.~(\ref{eq:BK2}). 
This structure can reverse the monotonic direction, but cannot produce an interior maximum.
By contrast, for three photons~($N=3$), 
cyclic three-photon permutations appear. 
The target-region and survival probabilities are written as 
$P_{\mathcal R}^{(T)}(x) = a_{\mathcal R}(1 + \beta_{\mathcal R}x^2 + \gamma_{\mathcal R}x^3)$, 
where ${\mathcal R} = {\mathcal K}$, ${\mathcal A}$, for $a_{\mathcal R}>0$. 
The coefficients are determined by $H_{\mathcal R}$ in Eq.~(\ref{eq:HR}), 
with $\beta_{\mathcal R}$ and $\gamma_{\mathcal R}$ representing 
the pairwise-exchange and cyclic three-photon coefficients
normalized by the distinguishable-particle contribution $a_{\mathcal R}$, respectively.
In Eq.~(\ref{eq:BK}), we assumed $a_{\mathcal A} = P_{\mathcal A}^{(T)}(0) > 0$. 
If $a_{\mathcal K}=0$, $H_{\mathcal K} \geq 0$ implies $P_{\mathcal K}^{(T)}(x)=0$ for all $x$, 
and hence $B_{\mathcal K}^{(T)}(x)=0$. 

For $a_{\mathcal K} > 0$, unlike the two-photon case in Eq.~(\ref{eq:dBK2}),
the numerator of the derivative retains an explicit $x$ dependence and can vanish for $0<x<1$,
depending on the relative coefficients for $\mathcal K$ and $\mathcal A$. 
More explicitly, the derivative of $B_{\mathcal K}^{(T)}(x)$ is given by
$a_{\mathcal A}a_{\mathcal K} x q(x)/(P_{\mathcal A}^{(T)}(x))^2$,
where
\begin{align}
  q(x) = 2 \Delta_2 + 3 \Delta_3 x + \Delta_{23} x^3, 
\label{eq:qx}
\end{align}
with 
$\Delta_2 = \beta_{\mathcal K} - \beta_{\mathcal A}$,
$\Delta_3 = \gamma_{\mathcal K} - \gamma_{\mathcal A}$, and 
$\Delta_{23} = \gamma_{\mathcal K}\beta_{\mathcal A} - \beta_{\mathcal K}\gamma_{\mathcal A}$. 

For $0 < x < 1$,
$q(x)$ determines the sign of the derivative and directly reveals the competition underlying the anomaly.
Since $q(0)=2\Delta_2$, the leading variation away from the fully distinguishable limit is governed solely
by the pairwise-exchange contrast $\Delta_2$.
For the circuit in Fig.~\ref{fig:example}, 
$\Delta_2>0$ is satisfied, so that $B_{\mathcal K}^{(T)}(x)$ increases for small $x$,
as shown in Fig.~\ref{fig:anomaly}(a).
By contrast, at the fully indistinguishable limit,
$q(1)=2\Delta_2+3\Delta_3+\Delta_{23} < 0$. 
The last two terms arise from the cyclic three-photon coefficients
and their coupling to the pairwise-exchange contribution. 
Thus, pairwise exchange drives the initial increase,
whereas the terms involving cyclic three-photon permutation
become sufficiently negative toward $x=1$ to reverse the trend.
The resulting sign change of $q(x)$ produces a maximum at partial distinguishability. 
The same mechanism can be viewed from the opposite direction. 
Reducing $x$ from unity suppresses the $x^3$-scaling cyclic contributions more rapidly 
than the $x^2$-scaling pairwise-exchange contributions.
Because the coefficients of these permutation-class contributions differ 
between the target-region and survival probabilities, 
this unequal suppression modifies the numerator and denominator of $B_{\mathcal K}^{(T)}(x)$
differently and can thereby increase their ratio. 
A necessary and sufficient condition for the anomaly is $\Delta_2>0$ and $q(1)<0$,
as proved in Appendix~\ref{appD}. 
The proof also shows that an interior minimum of $B_{\mathcal K}^{(T)}(x)$ is impossible.

 \section{Conclusion}
We have introduced a conditional generalized bunching probability for lossy multiphoton interference 
and shown that it can exhibit anomalous behavior for three photons. 
In the associated nonunitary linear optical interference process, 
partially distinguishable photons can be more strongly concentrated in a target region than fully indistinguishable photons. 
This demonstrates that, in a survival-conditioned ensemble, 
the internal loss changes the minimal hierarchy of generalized bunching.
Unlike nonlinear or measurement-induced photonic sampling schemes,
in which output statistics are modified through nonlinear interactions
or intermediate auxiliary measurements~\cite{Chabaud2021,Spagnolo2023,Francalanci2026},
the present effect arises in a passive linear optical circuit,
with conditioning applied only to the final survival events.
The minimal three-photon and three-mode setting considered here can be implemented with controlled internal loss,
tunable photon distinguishability, and photon-number-resolving detection up to three photons, 
placing an experimental demonstration within reach of current photonic technology.

Partial distinguishability has often been regarded as an imperfection in multiphoton interference. 
At the same time, it has increasingly been studied
as a useful degree of freedom in linear optical multiphoton interference~\cite{Seron2023,Chin2020,Jones2020}. 
Our results provide a perspective on exploiting partial distinguishability 
in survival-conditioned lossy multiphoton interference. 
For example, the presence or absence of conditional-bunching anomalies, 
together with their distinguishability-dependent profiles, 
could be useful for diagnosing the internal nonunitary interference structure of lossy optical circuits. 
More broadly, survival-conditioned statistics may offer a useful viewpoint 
for exploring multiphoton interference distributions
underlying lossy and partially distinguishable boson sampling~\cite{Aaronson2016,Garcia2019,Moylett2020}.
Finally, although we have focused on fixed single-photon inputs with pure internal states, 
extending the present analysis to other input-state features usually regarded as imperfections,
such as nontrivial photon statistics~\cite{Ikuta2026},
nonuniform Fock-state inputs~\cite{Shchesnovich2015,Khalid2018}, or internal-state mixedness~\cite{Jones2023},
may reveal unexplored interference regimes in lossy multiphoton systems. 

\begin{acknowledgments}
  We thank Tomohiro Yamazaki for helpful discussions and valuable comments.
  This work was supported by JST FOREST Program JPMJFR222V, 
  R \& D of ICT Priority Technology JPMI00316, and 
  JST Moonshot R \& D JPMJMS256K. 
\end{acknowledgments}

\appendix

\section{Nondecreasing generalized bunching for $N=3$, $M\geq 3$ and $|K|=1$}
\label{appA}

We prove that, for $N=3$, $M\geq 3$ and $|\mathcal K|=1$,
$B_{\mathcal K}^{(T)}(x)$ is a monotonically non-decreasing function of $x$ for $0\leq x\leq 1$. 
Since $H_{\mathcal K}$ is rank one, we obtain 
\begin{align}
  P_{\mathcal K}^{(T)}(x) = a_{\mathcal K}(1+3x^2+2x^3),
  \label{eq:PK3}
\end{align}
where $a_{\mathcal K}=(H_{\mathcal K})_{11}(H_{\mathcal K})_{22}(H_{\mathcal K})_{33}$.
On the other hand, $P_{\mathcal A}^{(T)}(x)$ is written as
\begin{align}
P_{\mathcal A}^{(T)}(x) = a_{\mathcal A}(1 + \beta_{\mathcal A}x^2 + \gamma_{\mathcal A}x^3), 
\end{align}
where $a_{\mathcal A} = (H_{\mathcal A})_{11}(H_{\mathcal A})_{22}(H_{\mathcal A})_{33}$, and 
\begin{align}
  \beta_{\mathcal A} &= |r_{12}|^2 + |r_{23}|^2 + |r_{31}|^2,
                       \label{eq:beta}\\
  \gamma_{\mathcal A} &= 2\U{Re}\left[ r_{12}r_{23}r_{31}\right],
                        \label{eq:gamma}
\end{align}
by introducing
\begin{align}
  r_{ii'} = \frac{(H_{\mathcal A})_{ii'}}{\sqrt{(H_{\mathcal A})_{ii}(H_{\mathcal A})_{i'i'}}}. 
\end{align}
If $a_{\mathcal K}=0$, Eq.~(\ref{eq:PK3}) gives $P_{\mathcal K}^{(T)}(x)=0$ for all $x$,
and hence $B_{\mathcal K}^{(T)}(x)=0$ due to $P_{\mathcal A}^{(T)}(x) > 0$.
Thus the monotonicity is trivial. 
In the following, we consider the case of $a_{\mathcal K} > 0$. 

Since $a_{\mathcal A} = P_{\mathcal A}^{(T)}(0) > 0$, 
the sign of the derivative of $B_{\mathcal K}^{(T)}(x)$ with respect to $x$ is the same as 
\begin{align}
  \od{}{x}\left( \frac{1+3x^2+2x^3}{1 + \beta_{\mathcal A} x^2 + \gamma_{\mathcal A} x^3}\right)
  = \frac{xf(x)}{(1 + \beta_{\mathcal A} x^2 + \gamma_{\mathcal A} x^3)^2}, 
\end{align}
where 
\begin{align}
f(x) = (6-2\beta_{\mathcal A}) + (6-3\gamma_{\mathcal A})x + (2\beta_{\mathcal A}-3\gamma_{\mathcal A})x^3.  
\end{align}
Thus, for $x > 0$, the sign of the derivative is determined by the sign of $f(x)$. 
For $x=0$, the derivative becomes zero. 

Since $H_{\mathcal A}$ is positive semidefinite, 
its principal $2\times 2$ submatrices are also positive semidefinite, which implies $|r_{ii'}|\leq 1$. 
Consequently, $\beta_{\mathcal A} \leq 3$ and $\gamma_{\mathcal A} \leq 2$ are satisfied. 
If $2\beta_{\mathcal A} - 3\gamma_{\mathcal A}\geq 0$, then $f'(x)\geq 0$. 
Together with $f(0)=6-2\beta_{\mathcal A}\geq 0$, this gives $f(x)\geq 0$ for $0 \leq x \leq 1$. 
If $2\beta_{\mathcal A} - 3\gamma_{\mathcal A} < 0$, then $f''(x) \leq 0$ for $0\leq x\leq 1$,
so $f(x)$ is concave on this interval. 
From $f(0)\geq 0$ and $f(1)=12-6\gamma_{\mathcal A}\geq 0$, the concavity gives $f(x)\geq 0$ 
for $0\leq x\leq 1$. 
Combining these results, the derivative of $B_{\mathcal K}^{(T)}(x)$ is non-negative, 
proving that $B_{\mathcal K}^{(T)}(x)$ is monotonically non-decreasing in $x$.

\section{Anomalous conditional generalized bunching induced by mode-dependent output attenuation}
\label{appB}
\begin{figure}
 \begin{center}
      \scalebox{0.7}{\includegraphics{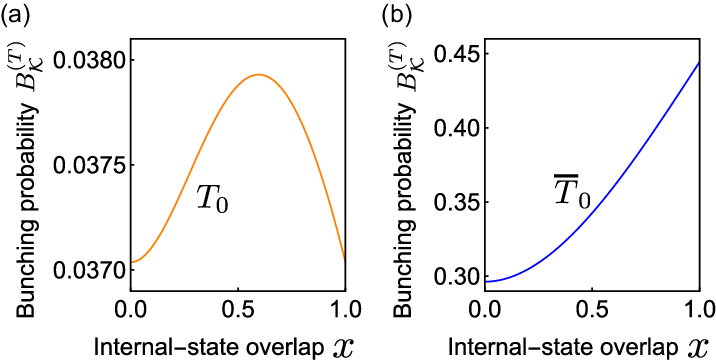}}
      \caption{
        (a) Conditional generalized bunching probability
        for the circuit $T_0=L^{(3)}F^{(3)}$ with $\eta_1=1/3$ and $\eta_2=1/6$, 
        showing a maximum at partial distinguishability. 
        (b) Corresponding probability for the row-normalized matrix $\overline{T}_0=F^{(3)}$,
        which increases monotonically with $x$. 
        The vertical scales are chosen independently in the two panels to make the variation with $x$ visible.
      }
 \label{fig:outputloss}
 \end{center}
\end{figure}
For $N=M=3$ and $\mathcal K=\{1,2\}$, 
we present an example of anomalous generalized bunching for a circuit described by $T=D_{\eta}U$.
We take $D_{\eta}=L^{(3)}=\U{diag}(\sqrt{\eta_1},\sqrt{\eta_2},1)$ and $U=F^{(3)}$,
where $F^{(3)}$ is the three-mode discrete Fourier transform.
Using the notation of the main text, this circuit corresponds to $T_0=L^{(3)}F^{(3)}$. 
Here, we choose output efficiencies different from those used for the three-photon example in the main text, $\eta_1=1/3$ and $\eta_2=1/6$. 
For this circuit, the conditional generalized bunching probability is $B_{\mathcal K}^{(T_0)}(x)=(1+x^2)/(3(9+7x^2+2x^3))$.
It attains its maximum $B_{\mathcal K}^{(T_0)}(x_{\max})\approx 0.0379$ at $x_{\max}\approx 0.596$,
whereas $B_{\mathcal K}^{(T_0)}(0) = B_{\mathcal K}^{(T_0)}(1)=1/27\approx 0.0370$.
The resulting $B_{\mathcal K}^{(T_0)}(x)$ is shown in Fig.~\ref{fig:outputloss}~(a).
The conditional generalized bunching probability clearly exhibits a maximum at partial distinguishability. 
In contrast, since $q_j=\eta_j$ in Eq.~(\ref{eq:qj}), the row-normalized matrix is $\overline{T}_0=F^{(3)}$. 
As shown in Fig.~\ref{fig:outputloss}~(b), $B_{\mathcal K}^{(\overline{T}_0)}(x)$ increases monotonically with $x$. 

This example illustrates an anomaly induced solely by mode-dependent output attenuation, which disappears under row normalization. 
By contrast, the anomaly discussed in the main text arises from the internal nonunitary interference structure 
and cannot be reduced to mode-dependent output attenuation.
As shown in Fig.~\ref{fig:anomaly}~(a), that anomaly persists after row normalization of the transfer matrix.

\section{Parameter regions for generalized bunching anomaly}
\label{appC}
\begin{figure}
 \begin{center}
      \scalebox{0.7}{\includegraphics{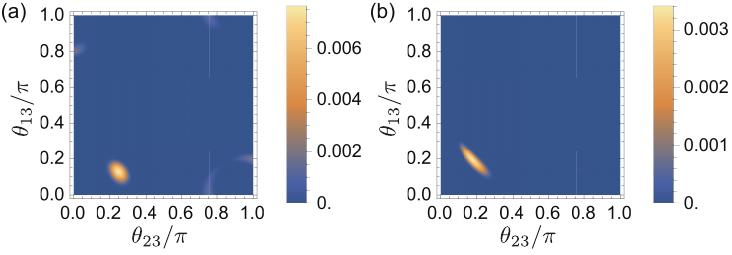}}
      \caption{
        Parameter regions for the generalized bunching anomaly in the three-photon circuit of the main text. 
        The color plots show $\Delta_{\mathcal K}^{(T)}$ 
        as a function of $\theta_{13}$ and $\theta_{23}$,
        with ${\mathcal K}=\{1,2\}$, $\eta_1=1/2$, and $\eta_2=1/4$. 
        Positive values of $\Delta_{\mathcal K}^{(T)}$ indicate that
        $B_{\mathcal K}^{(T)}(x)$ attains its maximum at partial distinguishability.
        (a) $\Delta_{\mathcal K}^{(T\u{int})}$.
        (b) $\Delta_{\mathcal K}^{(\overline{T}\u{int})}$.
      }
 \label{fig:delta}
 \end{center}
\end{figure}

We examine the robustness of the anomaly against variations of the circuit parameters.
For this purpose, we define
\begin{align}
\Delta_{\mathcal K}^{(T)}
        =
        \max_{0\leq x\leq 1} B_{\mathcal K}^{(T)}(x)
        -
        \max\{B_{\mathcal K}^{(T)}(0),\,B_{\mathcal K}^{(T)}(1)\}. 
\end{align}
Positive values of $\Delta_{\mathcal K}^{(T)}$ indicate that 
$B_{\mathcal K}^{(T)}(x)$ exceeds both endpoint values at some partial distinguishability~($0 < x < 1$). 
We evaluate $\Delta_{\mathcal K}^{(T)}$ as a function of $\theta_{13}$ and $\theta_{23}$,
with $\eta_1=1/2$, $\eta_2=1/4$, and ${\mathcal K}=\{ 1,2\}$, as shown in Figs.~\ref{fig:delta}~(a) and (b). 
Finite positive regions appear in both panels for $T\u{int}$ and $\overline{T}\u{int}$, 
demonstrating that the anomaly is not restricted to a fine-tuned parameter point.
Moreover, there are parameter regions where
$\Delta_{\mathcal K}^{(T\u{int})}>0$ and
$\Delta_{\mathcal K}^{(\overline{T}\u{int})}>0$ are simultaneously satisfied, 
which include the parameter set used in the main text,
$\theta_{13}=\pi/8$ and $\theta_{23}=\pi/4$.

\section{Necessary and sufficient condition for anomalous generalized bunching for $N=M=3$ and $|K|=2$}
\label{appD}
Before discussing the necessary and sufficient condition for the anomaly 
for $N=M=3$ and $|\mathcal K|=2$, we first consider the case $a_{\mathcal K}=0$.
For three photons, the target-region probability can be written as 
$P_{\mathcal K}^{(T)}(x) = a_{\mathcal K} + b_{\mathcal K}x^2 + c_{\mathcal K}x^3$, 
where
$a_{\mathcal K} = (H_{\mathcal K})_{11}(H_{\mathcal K})_{22}(H_{\mathcal K})_{33}$, 
$b_{\mathcal K} = (H_{\mathcal K})_{11}|(H_{\mathcal K})_{23}|^2+(H_{\mathcal K})_{22}|(H_{\mathcal K})_{31}|^2+(H_{\mathcal K})_{33}|(H_{\mathcal K})_{12}|^2$, 
$c_{\mathcal K} = 2\U{Re}\left[ (H_{\mathcal K})_{12}(H_{\mathcal K})_{23}(H_{\mathcal K})_{31}\right]$. 
Since $a_{\mathcal K}=0$, $(H_{\mathcal K})_{ii}=0$ for at least one $i$. 
The positive semidefiniteness of $H_{\mathcal K}$ gives
$(H_{\mathcal K})_{ii}(H_{\mathcal K})_{jj}\geq |(H_{\mathcal K})_{ij}|^2$, 
implying $|(H_{\mathcal K})_{ij}|^2=0$ for all $j$. 
It follows that $b_{\mathcal K}=c_{\mathcal K}=0$, and hence
$P_{\mathcal K}^{(T)}(x)=0$ for $0\leq x\leq1$.
Consequently, $B_{\mathcal K}^{(T)}(x)=0$ for $0\leq x\leq1$, and no anomaly occurs. 

In the following, we assume $a_{\mathcal K}>0$ and derive a necessary and sufficient condition 
for $B_{\mathcal K}^{(T)}(x)$ to attain a strict global maximum at some $x_*\in(0,1)$.
As shown in the main text,
the derivative of $B_{\mathcal K}^{(T)}(x)$ is given by
$a_{\mathcal A}a_{\mathcal K}xq(x)/(P_{\mathcal A}^{(T)}(x))^2$,
where $q(x)=2\Delta_2+3\Delta_3x+\Delta_{23}x^3$ in Eq.~(\ref{eq:qx}),
with
$\Delta_2=\beta_{\mathcal K}-\beta_{\mathcal A}$,
$\Delta_3=\gamma_{\mathcal K}-\gamma_{\mathcal A}$, and
$\Delta_{23}=\gamma_{\mathcal K}\beta_{\mathcal A}-\beta_{\mathcal K}\gamma_{\mathcal A}$.
Since $|\mathcal K|=2$, we have $\U{rank}(H_{\mathcal K})\leq 2$.
It follows that
$\U{det}(H_{\mathcal K})/a_{\mathcal K} =1-\beta_{\mathcal K}+\gamma_{\mathcal K} = 0$.
Using this relation, we obtain
\begin{align}
  q(x)=\Delta_2(2+3x+(1-d_{\mathcal A})x^3)-d_{\mathcal A}(3x+\beta_{\mathcal A}x^3), 
  \label{eq:qx2}
\end{align}
where $d_{\mathcal A}=1-\beta_{\mathcal A}+\gamma_{\mathcal A}$.
Since $H_{\mathcal A}$ is positive semidefinite,
$\U{det}(H_{\mathcal A})\geq0$, while Hadamard's inequality gives
$\U{det}(H_{\mathcal A})\leq a_{\mathcal A}$.
Therefore, $0 \leq d_{\mathcal A}=\U{det}(H_{\mathcal A})/a_{\mathcal A}\leq 1$.
Since $\beta_{\mathcal A}\geq 0$, if $\Delta_2\leq 0$, 
Eq.~(\ref{eq:qx2}) gives $q(x)\leq 0$ for $0\leq x\leq 1$.
Thus, $B_{\mathcal K}^{(T)}(x)$ is nonincreasing, and no strict interior maximum exists.
Hence, $\Delta_2 > 0$ is necessary for the anomaly.

We next suppose that $\Delta_2>0$ and that there exists an $x_*\in(0,1)$ such that $q(x_*)=0$. 
Since $d_{\mathcal A}=0$ would give $q(x) > 0$ from Eq.~(\ref{eq:qx2}),
the existence of the root requires $d_{\mathcal A}>0$.
Since 
\begin{align}
  q'(x)=3\Delta_2(1+(1-d_{\mathcal A})x^2)-3d_{\mathcal A}(1+\beta_{\mathcal A}x^2)
  \label{eq:dqx}
\end{align}
is linear in $x^2$ from Eq.~(\ref{eq:qx2}), 
it can be written as $q'(x)=(1-x^2)q'(0)+x^2q'(1)$. 
Solving $q(x_*)=0$ for $\Delta_2$ in Eq.~(\ref{eq:qx2}) 
and substituting the result into Eq.~(\ref{eq:dqx}), we obtain 
\begin{align}
  q'(0)
  &= -\frac{3d_{\mathcal A}\left(2-\gamma_{\mathcal A}x_*^3\right)}{2+3x_*+(1-d_{\mathcal A})x_*^3},\\
  q'(1)
  &= -\frac{3d_{\mathcal A}( 2(1+\beta_{\mathcal A})+\gamma_{\mathcal A}(3x_*-x_*^3))}{2+3x_*+(1-d_{\mathcal A})x_*^3}.
\end{align}
Given $d_{\mathcal A} > 0$, $q'(0) < 0$ follows from $\gamma_{\mathcal A}\leq2$ and $x_*<1$. 
$q'(1) < 0$ follows from $\gamma_{\mathcal A} > \beta_{\mathcal A}-1$, 
$\beta_{\mathcal A}\geq 0$, and $0 < 3x_*-x_*^3 < 2$. 
Thus, $q'(x) < 0$ is satisfied for all $0\leq x\leq 1$. 
This means that whenever an interior root exists,
$q(x)$ is strictly decreasing on $0\leq x\leq 1$, and the root is unique.

Since $q(0)=2\Delta_2>0$,
the condition $q(1)<0$, together with the continuity of $q(x)$,
guarantees the existence of an $x_*\in(0,1)$ such that $q(x_*) = 0$. 
From the above argument, this root is unique,
and $B_{\mathcal K}^{(T)}(x)$ attains its strict global maximum at $x=x_*$.
Conversely, if an interior root exists, the strict decrease of $q(x)$ implies $q(1) < q(x_*)=0$.
Hence, a necessary and sufficient condition for the anomaly to occur is that $\Delta_2>0$ and $q(1)<0$.
The above argument also shows that an interior minimum cannot occur.


\end{document}